\documentclass[aps,prl,twocolumn,preprintnumbers,10pt,showpacs]{revtex4-1}
\usepackage{amsmath,bm,amssymb}

\usepackage{graphicx}

\def\gapprox{\lower .7ex\hbox{$\;\stackrel{\textstyle >}{\sim}\;$}}

\allowdisplaybreaks

\begin{document}

\preprint{IPPP/16/38, ZU-TH 17/16, NSF-KITP-16-064}

\title{Precise QCD predictions for the production\\ of dijet final states in deep inelastic scattering}

\author{James Currie$^{a,c}$, Thomas Gehrmann$^{b,c}$, Jan Niehues$^b$}

\affiliation{
$^a$Institute for Particle Physics Phenomenology, Department of Physics, University of Durham, Durham, DH1 3LE, UK\\
$^b$Department of Physics, University of Z\"urich, Winterthurerstrasse 190, CH-8057 Z\"urich, Switzerland\\
$^c$Kavli Institute for Theoretical Physics, University of California,
Santa Barbara, CA 93106, USA
}


\begin{abstract}
The production of two-jet final states in deep inelastic scattering is an important QCD precision observable. We compute it 
for the first time to 
next-to-next-to-leading order (NNLO) in perturbative QCD. Our calculation is fully differential in the lepton and jet variables 
and allows one to impose cuts on the jets both in the laboratory and the Breit frame. We observe that the NNLO corrections 
are moderate in size, except at kinematical edges, and that their inclusion leads to a substantial reduction of the scale variation 
uncertainty on the predictions. Our results will enable the inclusion of deep inelastic dijet data in precision phenomenology 
studies. 
\end{abstract}

\maketitle

Our understanding of the inner structure of the proton has been shaped through a long series of deep-inelastic lepton-nucleon
experiments, which have established the partonic structure of the proton and provided precision measurements of 
parton distribution functions~\cite{disbook}. Specific combinations of the quark distributions can be probed in 
inclusive deep inelastic scattering (DIS), where the gluon distribution only enters indirectly as 
a correction and through scaling violations. A direct probe of the gluon distribution, which is 
less well constrained than the quark distributions, requires the selection of specific 
hadronic final states in deep inelastic scattering~\cite{Newman:2013ada}, such as heavy quarks or jets. 

Dijet final states in DIS are formed~\cite{lo} through the two basic
 scattering processes $\gamma^* q \to qg$ and $\gamma^* g \to q\bar q$, which vary in relative importance depending on 
 the kinematical region. Especially at low invariant masses of the dijet system, the gluon-induced process is 
 largely dominant. The interplay of lepton and dijet kinematics in this region allows the 
 gluon distribution to be probed over a substantial range.  The same process also provides 
 a direct measurement of the strong coupling constant $\alpha_s$. 
 
The DESY HERA electron-proton collider provided a large data set of hadronic final states in 
DIS at $\sqrt{s} = 319$~GeV. Dijet final states have been measured to high precision over a large kinematical range 
 by the H1~\cite{h11,h12,h1jet} and 
ZEUS~\cite{zeus1,zeus2}  experiments, that have also used these 
measurements in the determination of the strong coupling constant. 
The reconstruction of jets is performed in the Breit frame, defined by the 
direction of the virtual photon and incoming proton, while the jet rapidity coverage is limited by the  detector's 
geometry in the laboratory frame. Consequently, the definition of the fiducial phase space used in a jet measurement 
typically combines information from both frames. 

The interpretation of HERA data on  dijet production in DIS relies at present on theoretical 
predictions at next-to-leading order (NLO) 
in perturbative QCD~\cite{graudenz,mirkes,nagy}. The uncertainty associated with the NLO predictions
(as estimated through the variation of renormalization and factorization scales) is 
the main limitation to precision studies based on these data. In particular, they can not be 
included in a consistent manner in state-of-the-art determinations of parton 
distributions~\cite{abm,mmht,nnpdf,ct}, which typically require their 
input data to be described at next-to-next-to-leading order (NNLO) QCD accuracy.

In this letter, we present the first calculation of the next-to-next-to-leading order (NNLO)
 QCD prediction to dijet production in DIS. The QCD 
corrections at this order involve three types of scattering 
amplitudes:  the two-loop amplitudes for two-parton final states~\cite{Z3p2l},
the one-loop amplitudes for three-parton final states~\cite{Z4p1l} and the tree-level amplitudes for four-parton final 
states~\cite{Z5p0l}. The contribution from each partonic final state multiplicity contains infrared 
divergences from soft and collinear real radiation and from virtual particle loops; these 
infrared singularities cancel only once the different 
multiplicities are summed together for any infrared-safe final state definition~\cite{sw}. To implement the 
different contributions into a numerical program, a procedure for the extraction of all infrared  
singular configurations from each partonic multiplicity is needed. Several methods have been developed for this 
task at NNLO: sector decompostion~\cite{secdec}, $q_T$-subtraction~\cite{qtsub},
antenna subtraction~\cite{ourant}, sector-improved residue subtraction~\cite{stripper}, 
$N$-jettiness subtraction~\cite{njettiness} and colorful subtraction~\cite{trocsanyi}. 
\begin{figure*}[t]
\centering
\includegraphics[angle=0,width=0.3\linewidth]{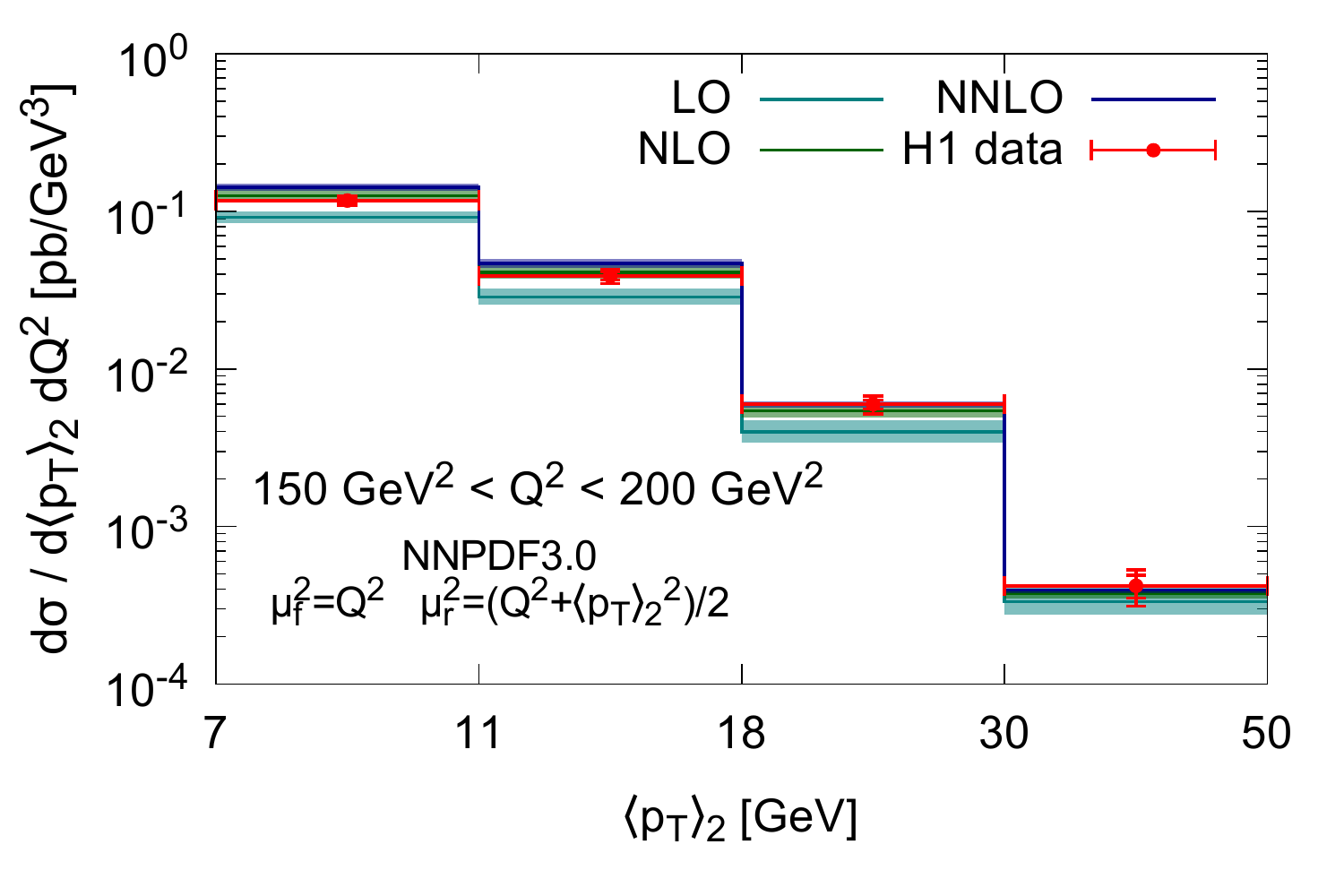}
\includegraphics[angle=0,width=0.3\linewidth]{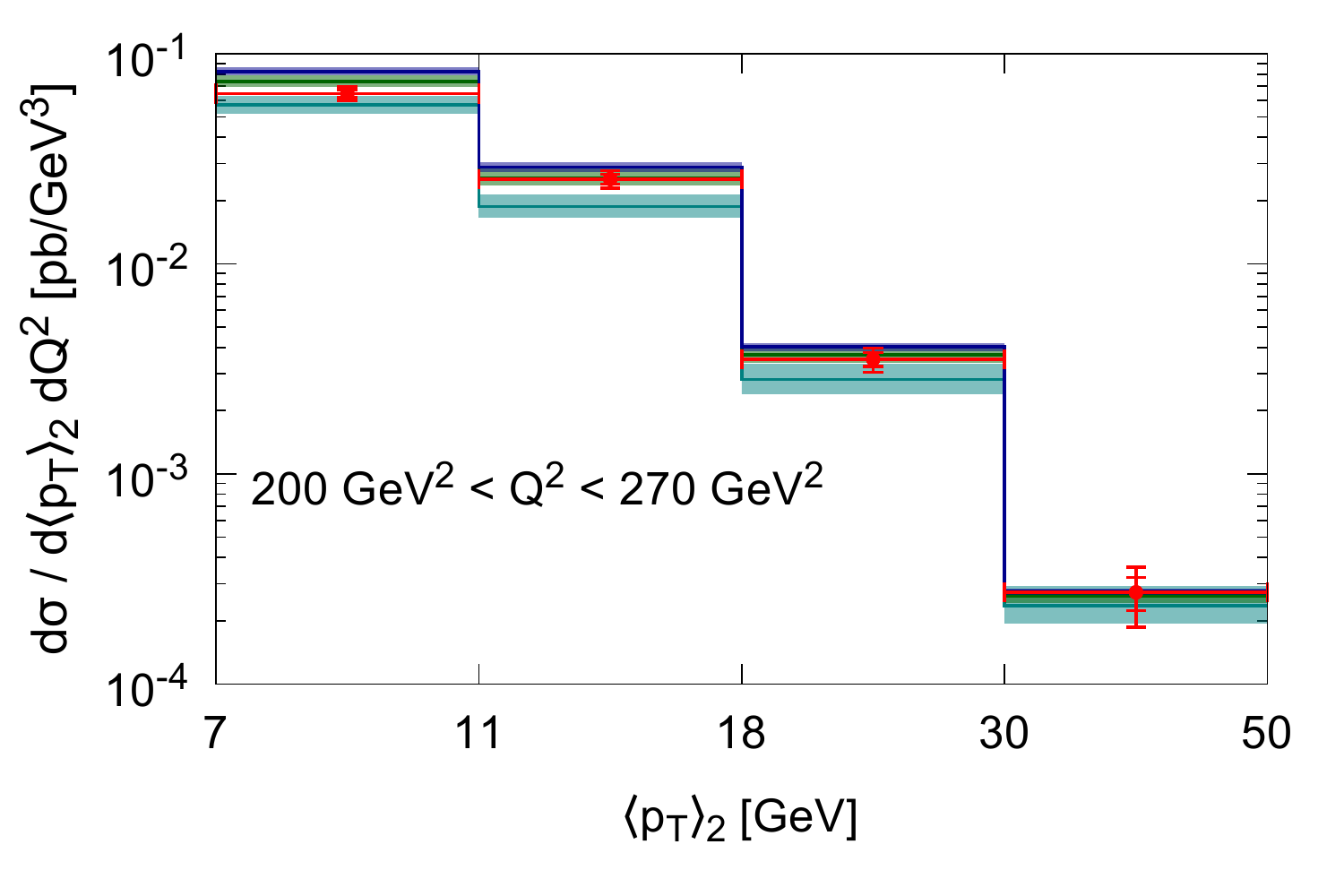}
\includegraphics[angle=0,width=0.3\linewidth]{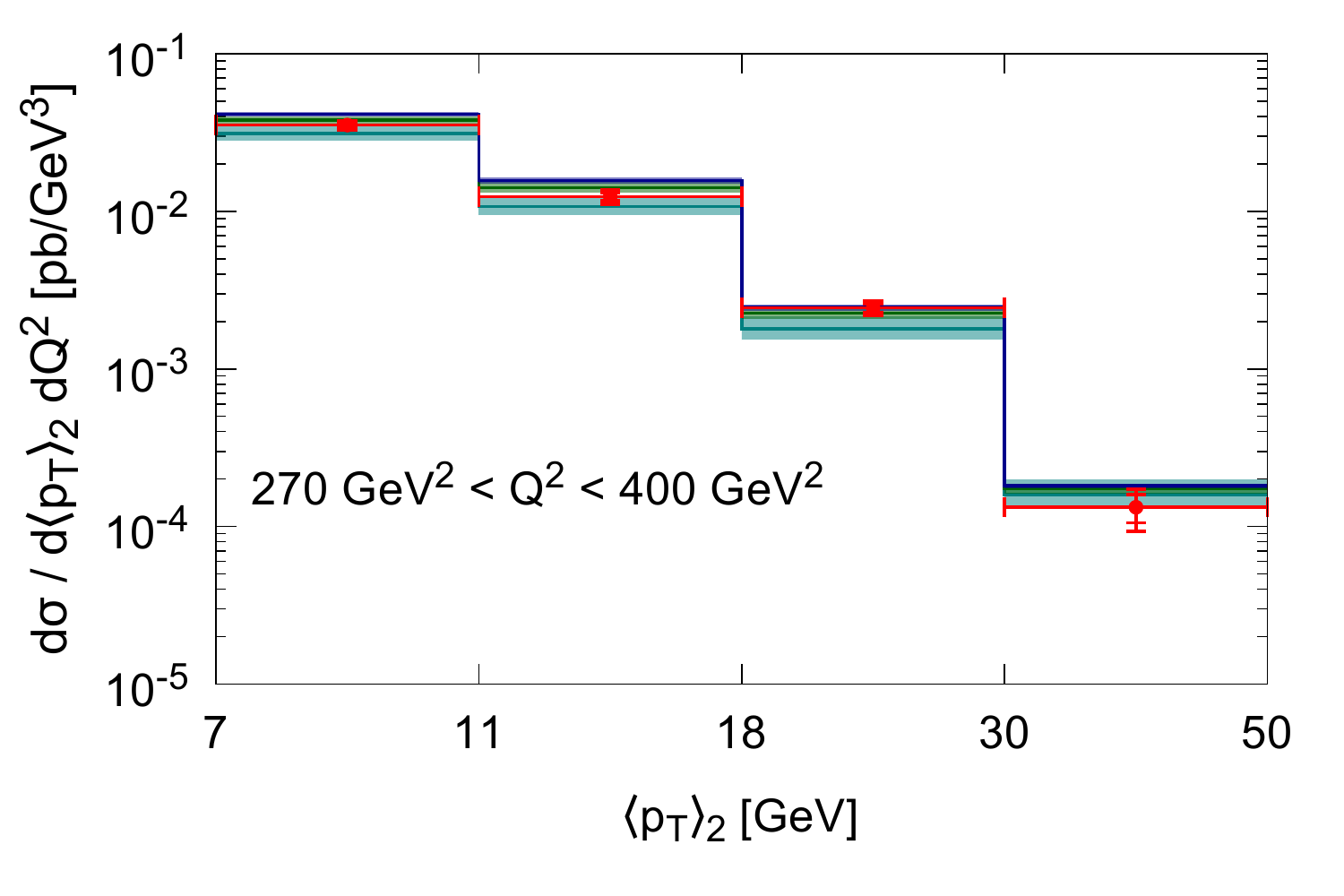}
\includegraphics[angle=0,width=0.3\linewidth]{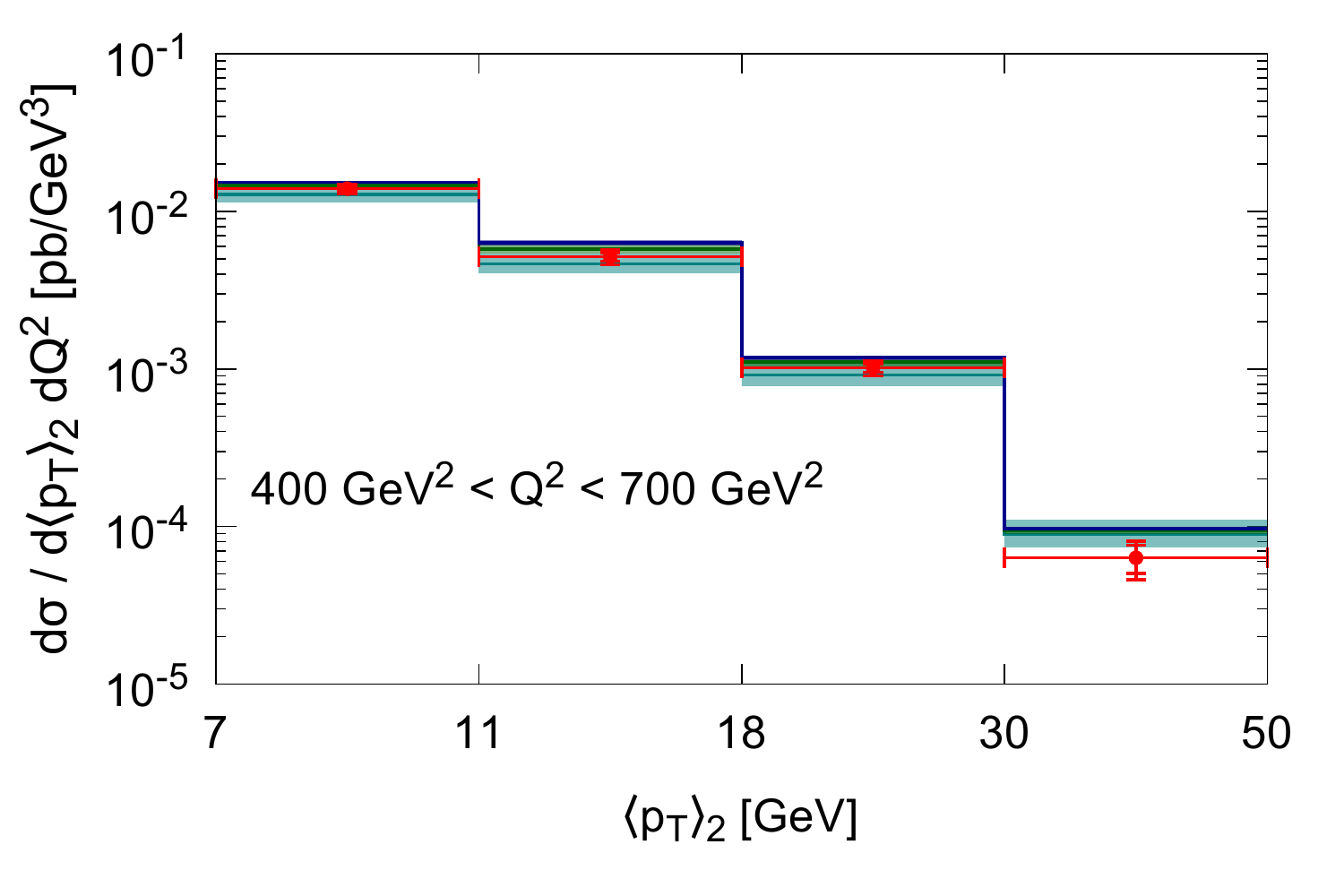}
\includegraphics[angle=0,width=0.3\linewidth]{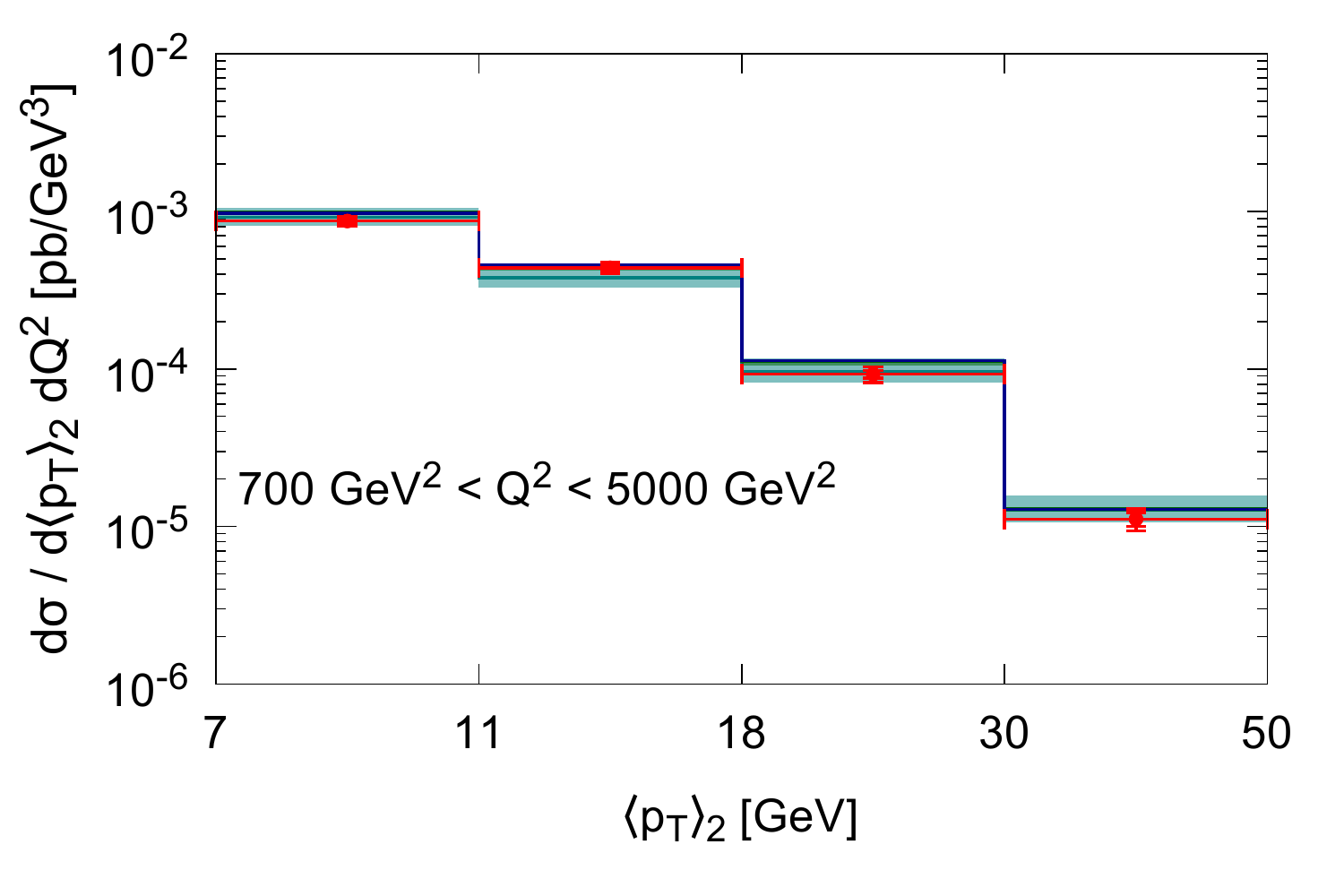}
\includegraphics[angle=0,width=0.3\linewidth]{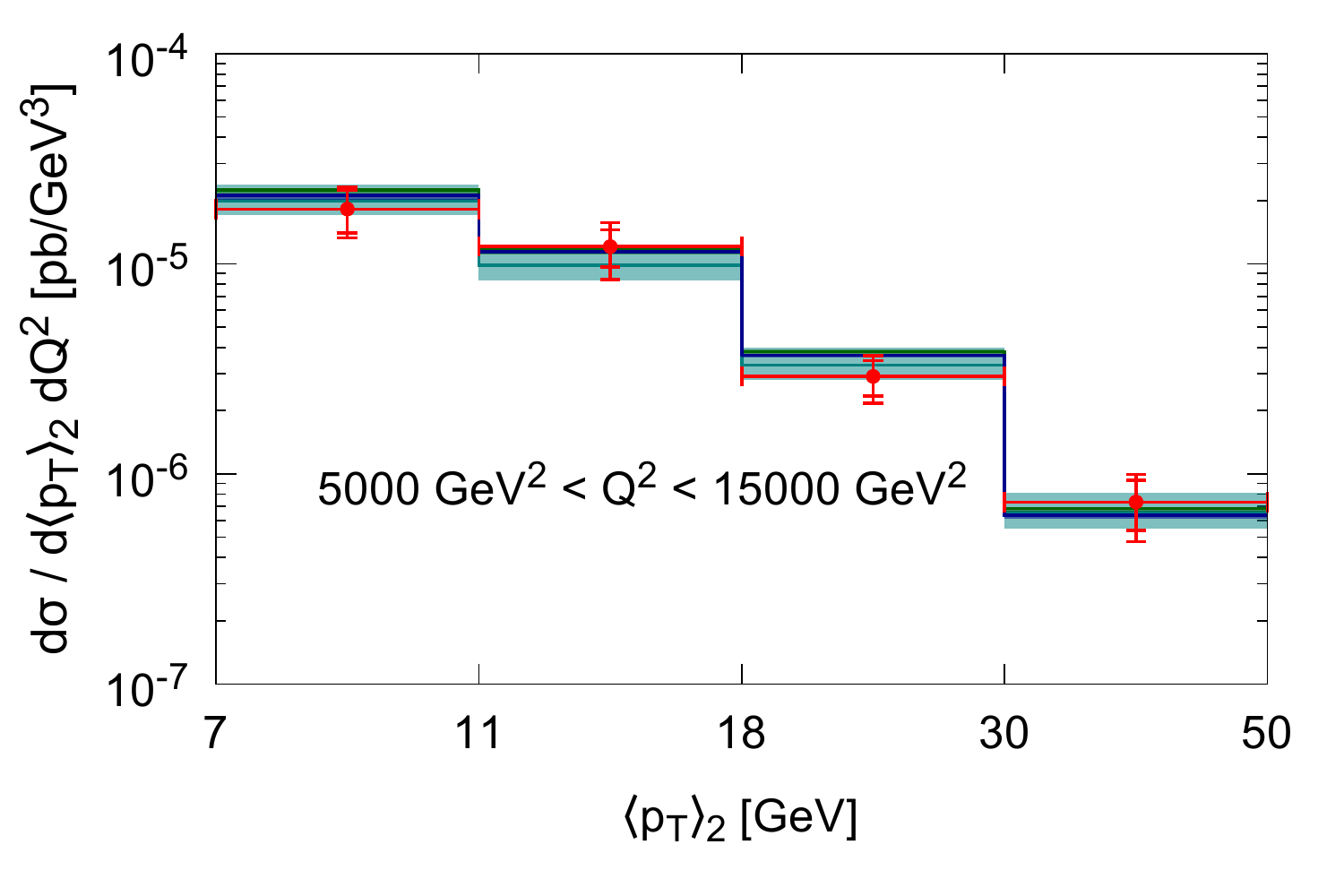}
\caption{
Inclusive dijet production in deep inelastic scattering as function of the average transverse momentum of the two leading 
jets in the Breit frame at LO, NLO, NNLO, compared to data from the H1 collaboration~\cite{h1jet}. \label{fig:ptdist}}
\end{figure*}

Our calculation is based on the antenna subtraction method~\cite{ourant}, which constructs the subtraction terms 
for the real radiation processes  out of antenna functions that encapsulate all color-ordered unresolved parton emission 
in between a pair of hard radiator partons, multiplied with reduced matrix elements of lower partonic multiplicity. 
By factorizing the final state phase space accordingly, it is possible to analytically integrate the antenna functions to 
make their infrared pole structure explicit, such that the integrated 
subtraction terms can be combined with the virtual corrections to 
yield a finite result. In the case of jet production in deep inelastic scattering, we need to use  
antenna functions with both hard radiators in the final state~\cite{ourant} and with one radiator in the initial and one 
in the final state~\cite{gionata}. 
The combination of real radiation contributions and unintegrated antenna subtraction terms is numerically finite in all 
infrared limits, such that all parton-level contributions to two-jet final states at NNLO can be implemented into a 
numerical program (parton-level event generator). This program can then incorporate the jet algorithm used in the 
experimental measurement as well as any type of event selection cuts.  A substantial part
 of the infrastructure of our program is common to other NNLO calculations of jet production observables 
 within the antenna subtraction method~\cite{eerad3,nnlo2j,nnlott,nnlohj,nnlozj}, which are all  
 part of a newly developed code named NNLOJET. To validate our 
 implementation of the tree-level and one-loop matrix elements, 
 we compared the NLO predictions for dijet and trijet production against SHERPA~\cite{sherpa} (in 
 DIS kinematics~\cite{hoeche}), which
uses OpenLoops~\cite{openloops} to automatically generate the  one-loop contributions at NLO. The antenna 
subtraction is then verified by testing the convergence of subtraction terms and matrix elements in all unresolved limits 
(as documented for example in~\cite{joao}) and by the infrared pole cancellation between the integrated subtraction terms 
and the two-loop matrix elements.

As a first application of our calculation, we consider the recent measurement by the H1 collaboration~\cite{h1jet} 
of dijet production in DIS at high virtuality $Q^2$. The measurement was performed on data taken at the 
DESY HERA electron proton collider at a centre-of-mass energy of $\sqrt{s} = 319$~GeV. Deep inelastic scattering events 
are selected by requiring the range of lepton scattering variables: 
exchanged boson virtuality 150~GeV$^2<Q^2<$15000~GeV$^2$ and energy transfer in 
the proton rest system $0.2<y<0.7$. The hadronic final state is boosted to the Breit frame of reference, where the 
jet clustering is performed using the inclusive hadronic $k_T$ algorithm~\cite{hadkt} with $E_T$ recombination. 
To ensure that the jets are
contained in the calorimeter coverage, a cut on their pseudorapidity is applied in the HERA laboratory frame:
$-1.0<\eta_{L}<2.5$. Jets are accepted in the inclusive dijet sample if their transverse momentum in the 
Breit frame is  5~GeV$<p_{T,B}<50$~GeV and are ordered in this variable. The event is retained if 
the invariant mass of the two leading jets 
is $M_{12}>16$~GeV. The H1 collaboration provides double differential distributions in $Q^2$ and either 
the average transverse momentum 
of the two leading jets $\langle p_T\rangle_2= (p_{T1,B} + p_{T2,B})/2$ or the variable $\xi_2 = x (1+M_{12}/Q^2)$ where 
$x$ is the Bjorken variable reconstructed from the lepton kinematics. At leading order, $\xi_2$   can be identified with the 
proton momentum fraction carried by the parton that initiated the hard scattering process. 

The theoretical predictions use the NNPDF3.0 parton distribution functions~\cite{nnpdf} with $\alpha_s(M_Z^2) = 0.118$ and 
are evaluated with
default renormalization and factorization scales  $\mu_R = \sqrt{Q^2+\langle p_T\rangle_2^2}$, $\mu_F = \sqrt{Q^2}$. The uncertainty on the theoretical 
prediction from missing higher orders is 
estimated by varying these scales by a factor between $1/2$ and $2$. The electromagnetic coupling is also evaluated at 
a dynamical scale as $\alpha(Q^2)$ according to QED evolution, with $\alpha(100~\mbox{GeV}^2)=0.0075683$. 
The theoretical predictions are corrected bin-by-bin for hadronization and electroweak effects using the
tables provided in~\cite{h1jet}. 
\begin{figure*}[t]
\centering
\includegraphics[angle=0,width=0.3\linewidth]{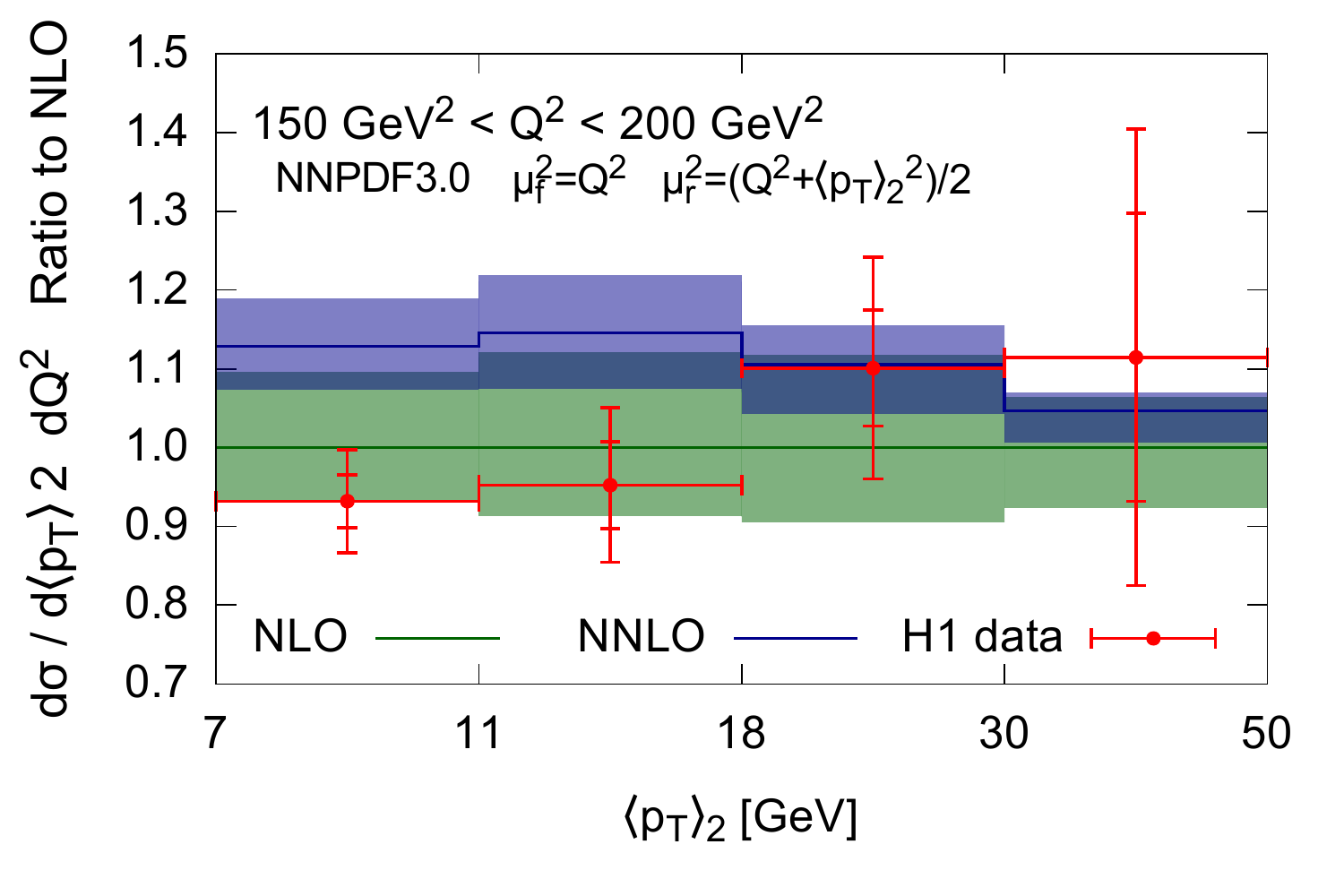}
\includegraphics[angle=0,width=0.3\linewidth]{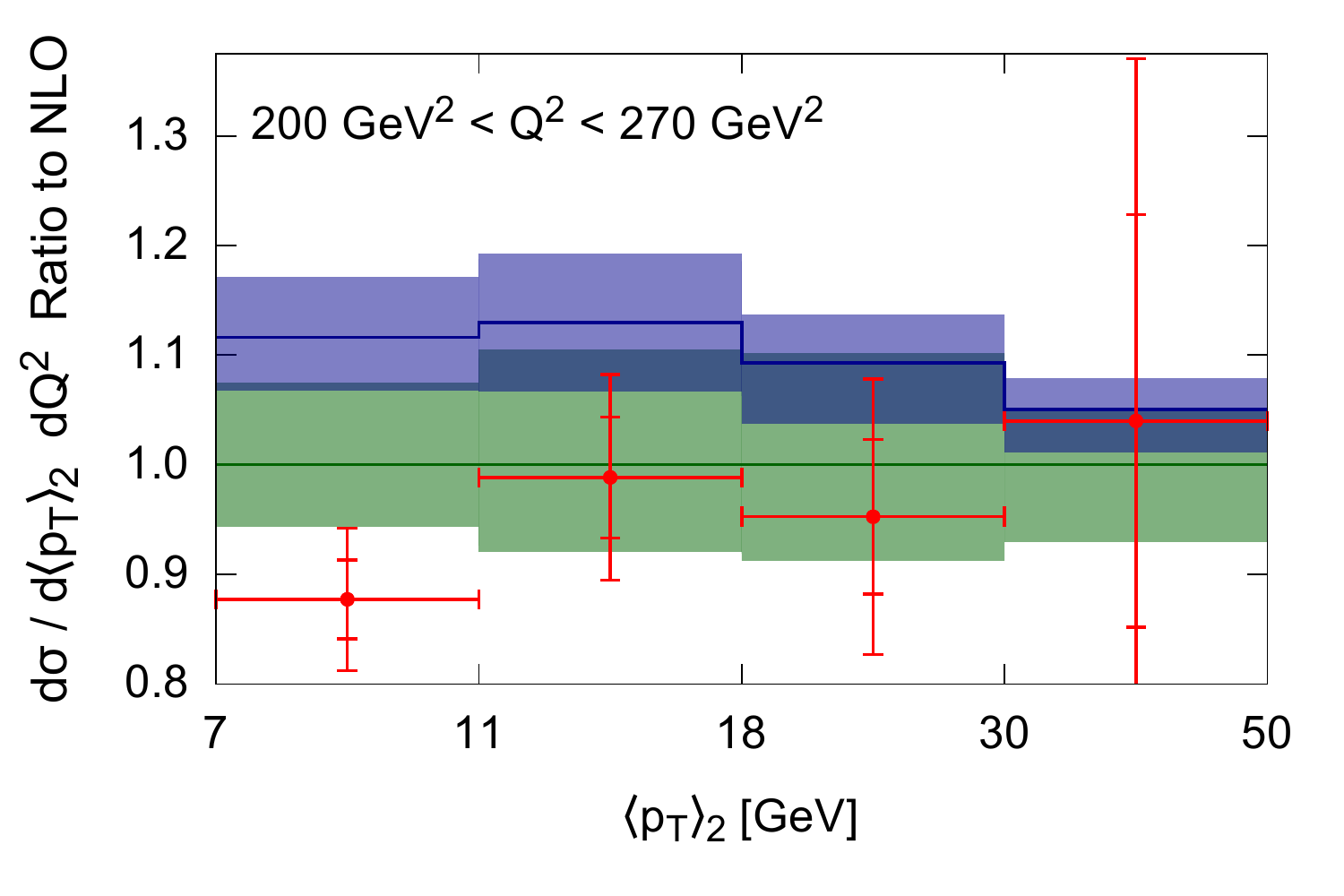}
\includegraphics[angle=0,width=0.3\linewidth]{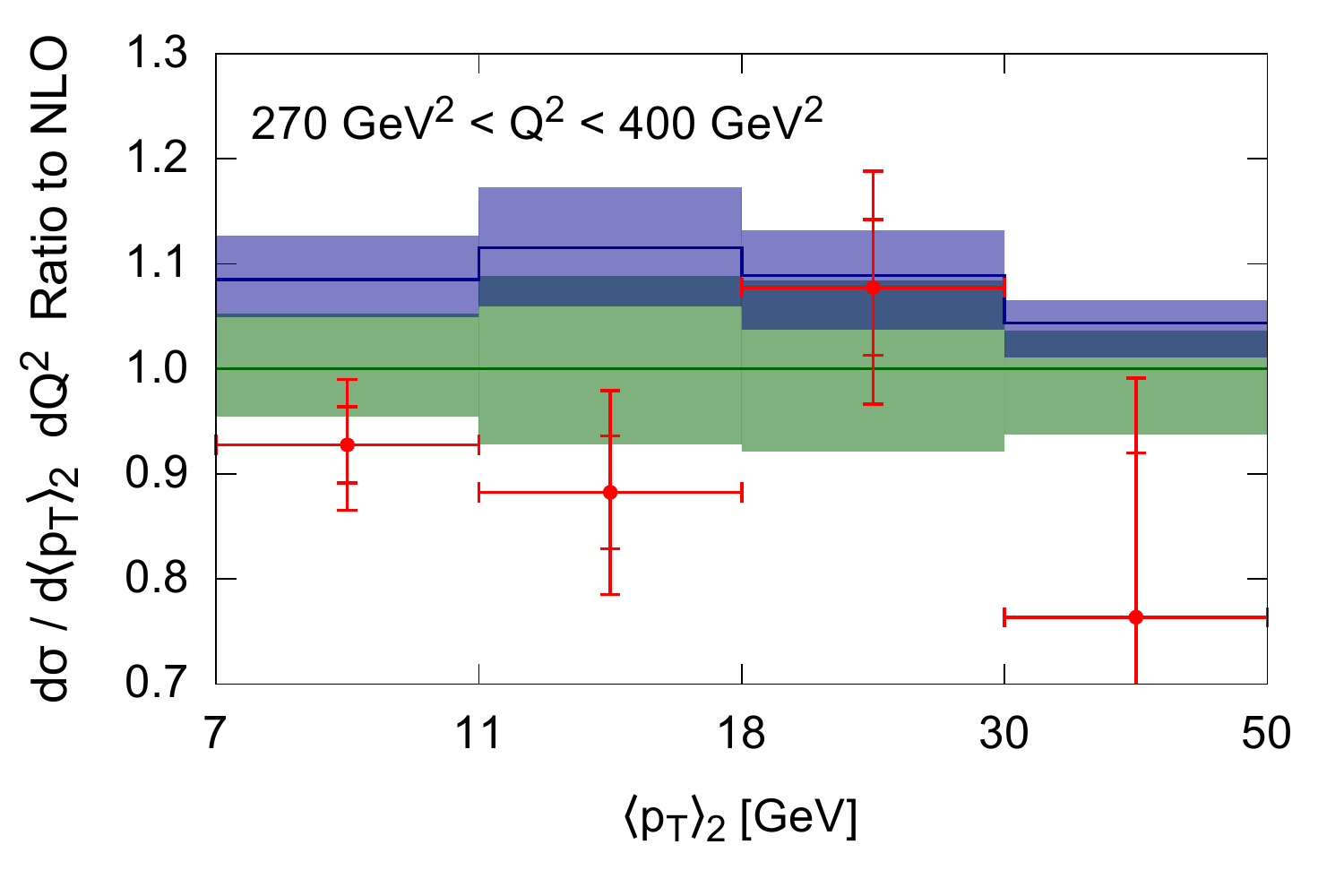}
\includegraphics[angle=0,width=0.3\linewidth]{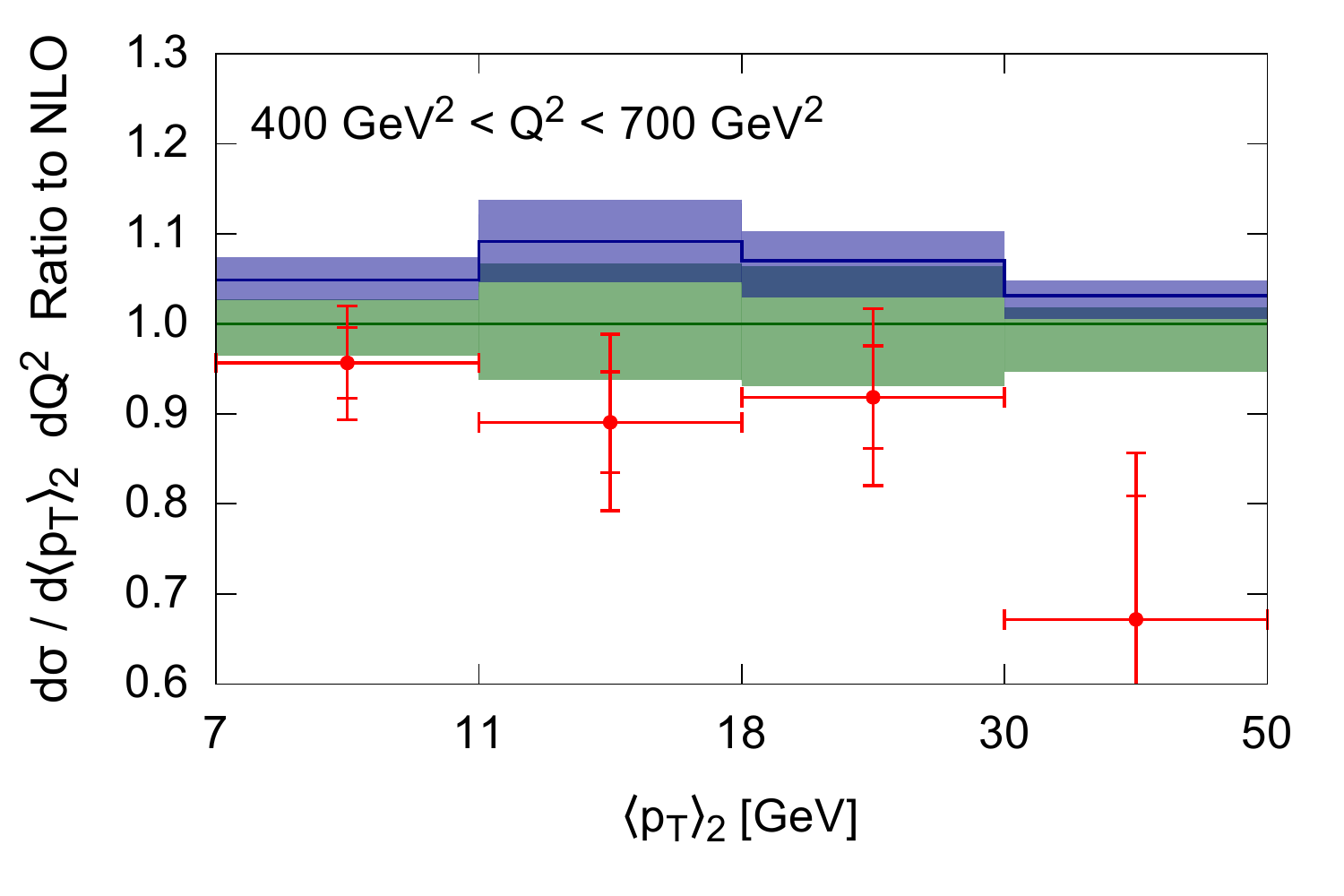}
\includegraphics[angle=0,width=0.3\linewidth]{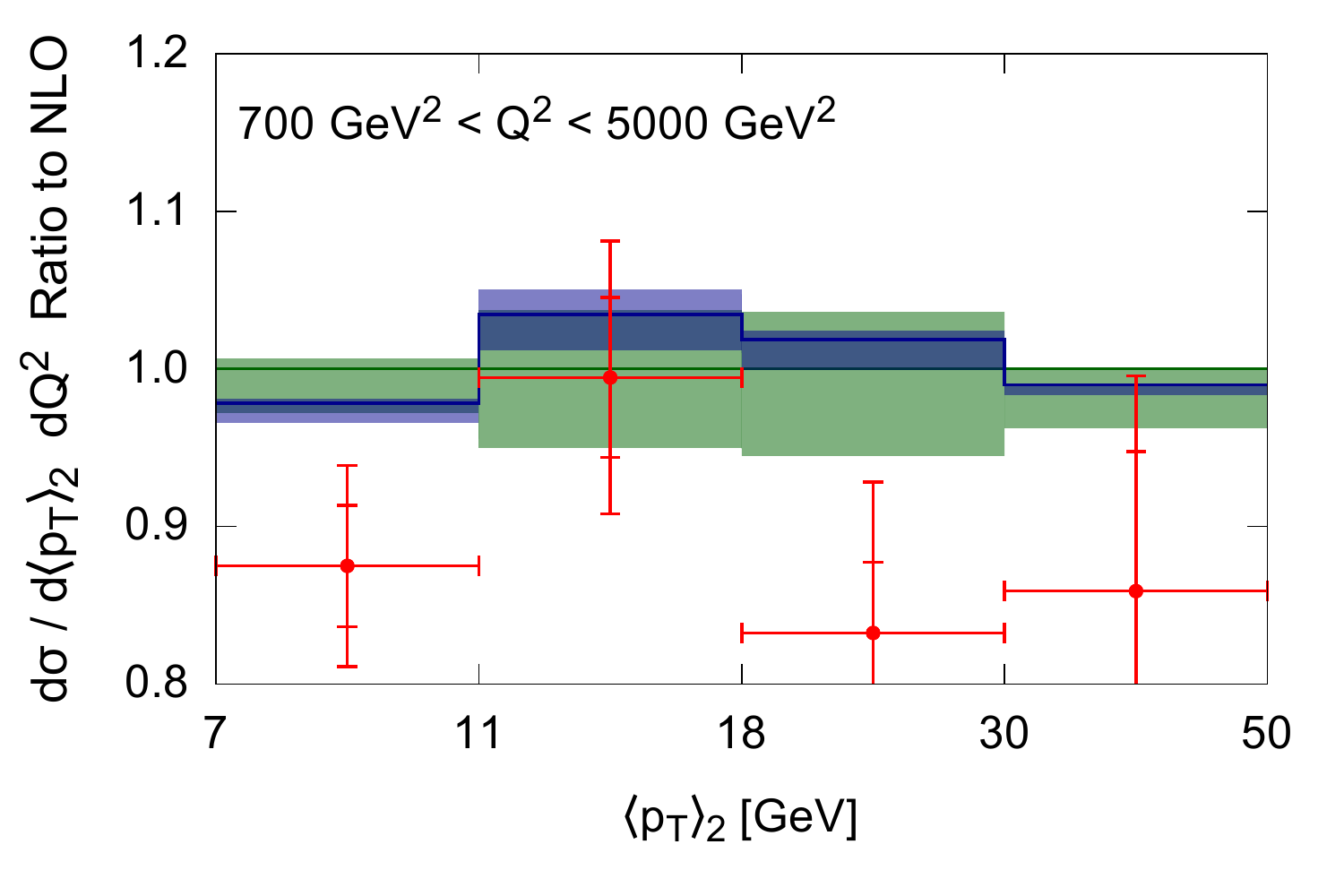}
\includegraphics[angle=0,width=0.3\linewidth]{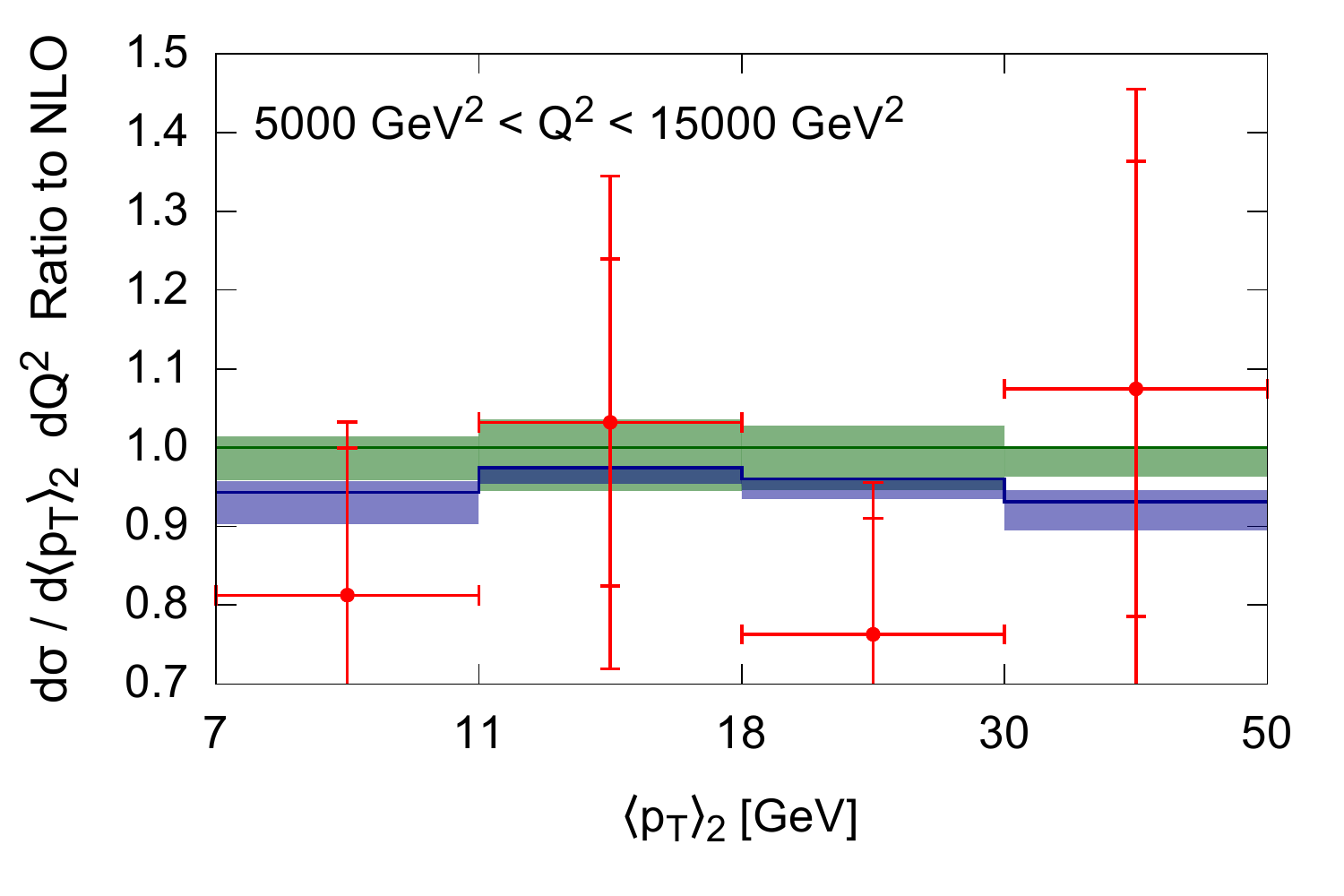}
\caption{
Inclusive dijet production in deep inelastic scattering as function of the average transverse momentum of the two leading 
jets in the Breit frame normalized to NLO and compared to data from the H1 collaboration~\cite{h1jet}. \label{fig:ptratio}}
\end{figure*}

Figure~\ref{fig:ptdist} displays the $\langle p_T\rangle_2$ distribution in six $Q^2$ bins. For better visibility, the same plots are 
normalized to the NLO prediction in Figure~\ref{fig:ptratio}, excluding the LO contribution which is typically considerably below the 
NLO curve and is associated with a large error. We observe that for all but the first bins in $\langle p_T\rangle_2$, the NNLO 
predictions are inside the NLO uncertainty band and that their inclusion leads to a substantial reduction of the 
theory uncertainty to typically 5\% or less (especially at high $Q^2$), which is now below 
the statistical and systematical uncertainty on the 
experimental data. We observe that the theoretical NNLO predictions tend to be above the experimental data. This 
feature points to the potential impact that the inclusion of these data could have in a global determination of 
parton distributions and of the strong coupling constant at NNLO 
accuracy. The tension between data and NNLO predictions is largest at lower values of $Q^2$, where the data is most accurate and 
the gluon-induced 
subprocess dominates the dijet production cross section. 

The first bins in $\langle p_T\rangle_2$ display a larger correction, often at the upper boundary of the NLO band, and only a mild reduction in 
scale uncertainty. They already have very large NLO corrections, typically with a NLO/LO ratio of about 2. This feature can be 
understood from a sophisticated interplay of the  $M_{12}>16$~GeV cut with the other jet cuts. The  $M_{12}$ cut 
forbids a substantial part of the phase space relevant to the first bin in the $\langle p_T\rangle_2$ distribution to be 
filled by the leading order process. This results in a perturbative instability~\cite{sudakov} 
starting below $\langle p_T\rangle_2=8$~GeV, which leads to a destabilization of the perturbative series for the first bin. 

\begin{figure}[b]
\centering
\includegraphics[angle=0,width=0.7\linewidth]{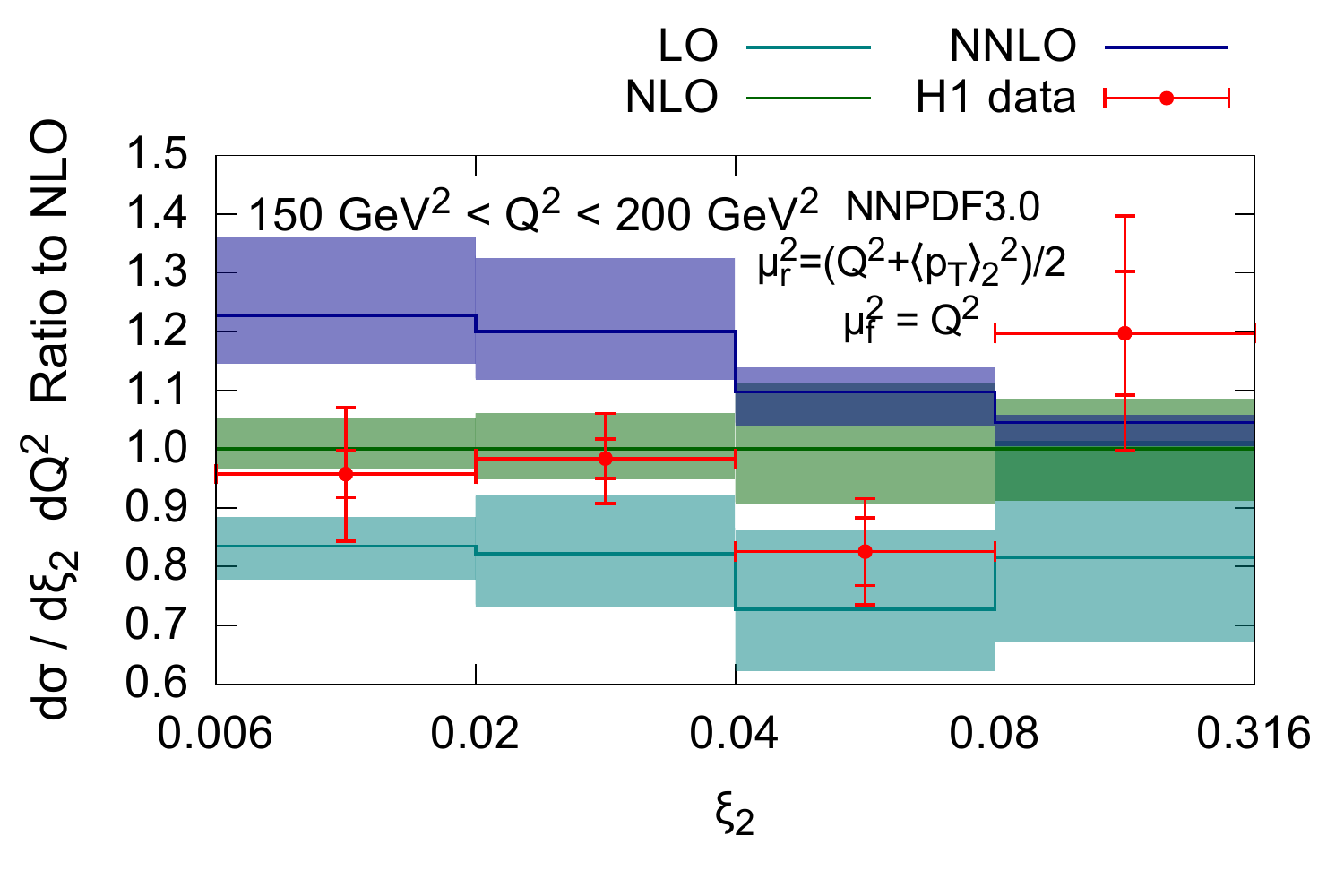}
\caption{
Inclusive dijet production in deep inelastic scattering as function of $\xi_2$ normalized to NLO and compared to data from the H1 collaboration~\cite{h1jet}. \label{fig:xiratio}}
\end{figure}

To further illustrate this issue, we display the $\xi_2$ distribution in the lowest bin in $Q^2$ in 
Figure~\ref{fig:xiratio}. 
The same perturbative instability is present, now spread more 
uniformly over the first two bins. It is more pronounced than in the $\langle p_T\rangle_2$ distribution 
due to the fact that 
an even larger fraction of the phase space is forbidden at leading order, since jets down to $p_{T,B}=5$~GeV are 
accepted in this distribution, while maintaining the  $M_{12}>16$~GeV cut. The resulting instability can already be 
seen in going from LO to NLO, with substantial corrections outside the nominal scale variation band. 
In the bins with larger $\xi_2$, events with low $M_{12}$ close to the cut are of lower importance, resulting in a better 
perturbative convergence and a more reliable prediction.   

In this letter, we presented the first calculation of dijet production in deep inelastic scattering to NNLO in QCD. Our results are
fully differential in the kinematical variables of the final state lepton and the jets. We applied our calculation to 
the kinematical situation that is relevant to a recent dijet measurement by the H1 collaboration~\cite{h1jet}. 
Except for jet production at low transverse momentum (where the experimental event selection 
cuts destabilize the perturbative convergence), we observe the NNLO corrections to be moderate in size, and 
overlapping 
with the scale uncertainty band of the previously available NLO calculation. Especially at lower $Q^2$, the NNLO predictions 
tend to be above the data, which could provide important new information on the gluon distribution at NNLO. 
The residual uncertainty on the NNLO results 
is of the order of 5\% or less, and below the errors on the experimental data. Our results enable 
the inclusion of deep inelastic jet data into precision phenomenology studies of the structure of the proton and of the strong 
coupling constant. 

We would like to thank Nigel Glover, Alexander Huss and Thomas Morgan for many 
interesting discussions throughout the whole course of this project, Stefan H\"oche and 
Marek Sch\"onherr for help with the NLO comparisons 
against SHERPA and Daniel Britzger for
useful clarifications on the 
H1 jet data. 
This research was supported in part by the Swiss National Science Foundation (SNF) under contract 200020-162487, 
 in part by
the UK Science and Technology Facilities Council as well as by the Research Executive Agency (REA) of the European Union under the Grant Agreement PITN-GA-2012-316704  (``HiggsTools''),  the ERC Advanced Grant MC@NNLO (340983) and
 by the National Science Foundation under grant NSF PHY11-25915.

\end{document}